\documentclass[prd,nofootinbib]{revtex4}
\usepackage{epsfig,amsmath,amssymb}
\usepackage{dcolumn}% Align table columns on decimal point
\usepackage{bm}% bold math
\def\d{{\rm d}}

\newcommand{\e}{{\rm e}}

\newcommand{\omegav}{\boldsymbol{\omega}}
\newcommand{\Omegav}{\boldsymbol{\Omega}}

\newcommand{\Piv}{\boldsymbol{\Pi}}
\newcommand{\piv}{\boldsymbol{\pi}}
\newcommand{\p}{{\rm p}}
\newcommand{\x}{{\rm x}}
\begin{document}

\title{Angular momentum conservation in heavy ion collisions at very high energy}
\author{F. Becattini}
\email{francesco.becattini@fi.infn.it}
\affiliation{Dipartimento di Fisica Universit\`{a} di Firenze and INFN Sezione di 
Firenze, Florence, Italy}
\author{F. Piccinini}
\email{fulvio.piccinini@pv.infn.it}
\affiliation{INFN, Sezione di Pavia, Pavia, Italy}
\author{J. Rizzo}
\email{rizzoj@lns.infn.it}
\affiliation{Dipartimento di Fisica Universit\`{a} di Firenze, Florence, Italy}

\date{November 8 2007}% It is always \today, today,
             %  but any date may be explicitly specified

\begin{abstract}
The effects of angular momentum conservation in peripheral heavy ion collisions 
at very high energy are investigated. It is shown that the initial angular 
momentum of the quark-gluon plasma should enhance the azimuthal anisotropy of particle 
spectra (elliptic flow) with respect to the usual picture where only the 
initial geometrical eccentricity of the nuclear overlap region is responsible for 
the anisotropy. In hydrodynamical terms, the initial angular momentum entails a
non trivial dependence of the initial longitudinal flow velocity on the transverse
coordinates. This gives rise to a non-vanishing vorticity in the equations of
motion which enhances the expansion rate of the supposedly created fluid compensating 
for the possible quenching effect of viscosity.
A distinctive signature of the vorticity in the plasma is the generation of an 
average polarization of the emitted hadrons, for which we provide analytical 
expressions. These phenomena might be better observed at LHC, where the initial 
angular momentum density will be larger and where we envisage an increase of the 
elliptic flow coefficient $v_2$ with respect to RHIC energies.
\end{abstract}

\maketitle

%*****************************************************************************
\section{Introduction}
%*****************************************************************************

Nuclei colliding at ultrarelativistic energies have a large initial orbital
angular momentum $L_0$ if their impact parameter is of order of some fm; in fact,
for symmetric nuclei, $L_0 \simeq A \sqrt{s}_{NN} b/2$ in natural units ($\hbar = 1$).
For Au-Au collisions at RHIC energies $\sqrt{s}_{NN}=200$ GeV and $L_0 \sim 5 \times 10^5$ 
at an impact parameter $b=5$ fm. The angular momentum will be almost two order of 
magnitude larger in the forthcoming Pb-Pb collisions at LHC, at $\sqrt{s}_{NN}=5.5$ 
TeV, with $L_0 \sim 1.4 \times 10^7$. Due to the inhomogeneity of the colliding nuclei in 
the transverse plane, a significant fraction of $L_0$ {\em must} be deposited in 
the interaction region, in other words should be transferred to the supposedly formed
Quark-Gluon Plasma (QGP). 
Large values of the initial angular momentum of the plasma may give rise, as we will 
show, to significant observables effects. 

According to the to-date generally accepted description of the collision process, 
a locally equilibrated plasma is formed after a relatively short proper time 
(of the order of 1 fm/$c$) followed by a purely ideal-fluid hydrodynamical expansion. 
This kind of approach proved to be able to reproduce the large observed values of
the elliptic flow in peripheral collisions, at a finite impact parameter, and
the transverse momentum spectra of particles in the low $p_T$ region \cite{review}. 
Usually, in this kind of description, the Bjorken hydrodynamics scaling hypothesis
is used either all along the evolution (2+1 hydro) \cite{heinz} or just at the initial 
proper time (3+1 hydro \cite{hirano}). In both cases, the initial longitudinal 
flow velocity only depends on $z$ which amounts to make the initial angular momentum
vanishing unless the energy density has an asymmetric dependence on the transverse
coordinates \cite{hirano}. But even if this was assumed, and the initial angular 
momentum was then non-vanishing, the dynamical evolution would be different from
the case of a longitudinal flow velocity depending on transverse coordinates, as
we will show later.

In recent papers \cite{romat,hsong} it has been found that amending the ideal
fluid assumption with even a minimal viscosity strongly affects the elliptic flow.
Particularly, Heinz and Song pointed out that in order to restore the agreement 
with a hydrodynamical description, one should enforce significant modifications 
of the initial conditions or the equation of state, such that the authors 
raise some doubts about the interpretation of RHIC results. In this paper, we want 
to show that including the initial angular momentum by a suitable modifications of 
the initial fluid velocity profile may cure the problem, or at least it may give a 
contribution in this direction. In fact, a finite angular momentum enhances the 
elliptic flow coefficient and broadens the transverse momentum spectra, exactly what 
is needed to counterbalance the quenching effect of viscosity.

The most distinctive signature of an intrinsic angular momentum would be the polarization
of the emitted hadrons. This argument has been put forward in refs.~\cite{xnwang,wang}
where the authors take a QCD perturbative approach. Also, more recently, 
polarization has been related to the fluid vorticity \cite{torrieri}, yet without 
developing an explicit mathematical relation. In this paper, we take advantage
of a very recent study of the ideal relativistic spinning gas \cite{becapicc} and
present a formula relating polarization to the angular velocity of an equilibrated,
i.e. rigidly rotating, hydrodynamical system. We argue, on the basis of the 
locality principle, that such formula should hold for the most general fluid 
motion where the angular velocity is to be presumably replaced by an expression
involving the local acceleration, hence the vorticity, of the fluid.

%-------------------------------------------------------------------------
\begin{figure}
  \includegraphics[height=.4\textheight]{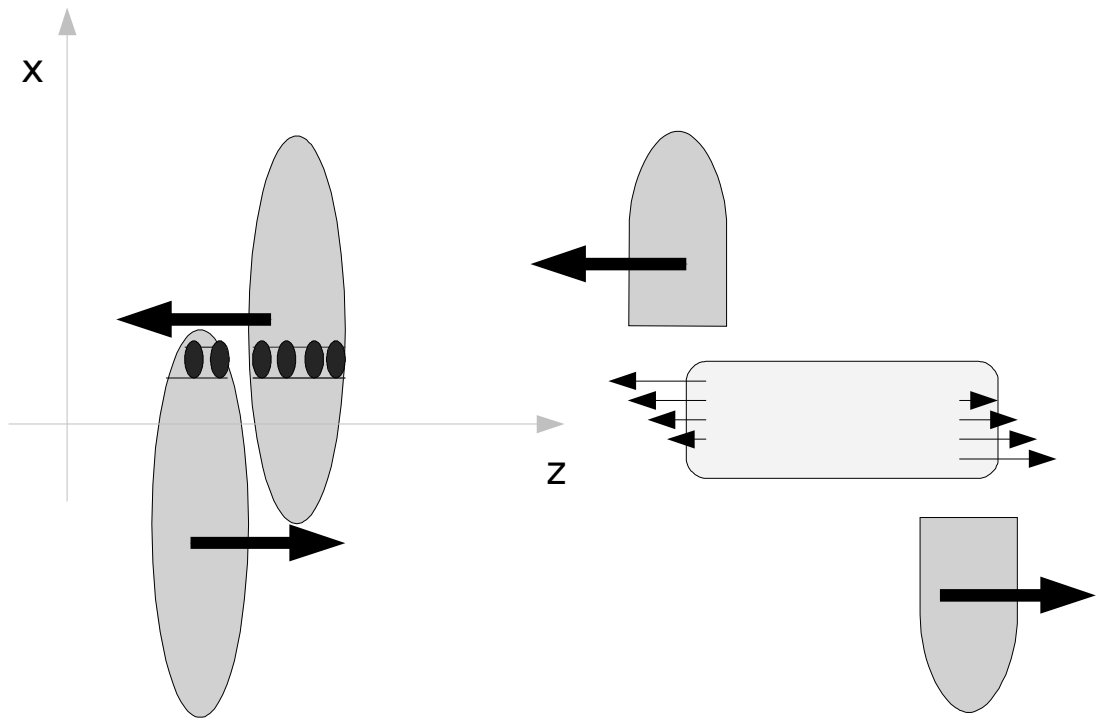}
  \caption{Sketch of a peripheral heavy ion collision at very high energy in
           the longitudinal projection. The initial momentum distribution of the
           interaction region (right) should have a gradient along the axis $x$ 
           orthogonal to the collision axis $z$ stemming from the different 
           transverse densities of the colliding strips (left).}
\label{collision}
\end{figure}
%-------------------------------------------------------------------------

%*****************************************************************************
\section{Angular momentum conservation in heavy ion collisions}
%*****************************************************************************

%-------------------------------------------------------------------------
\begin{figure}
  \includegraphics[height=.4\textheight]{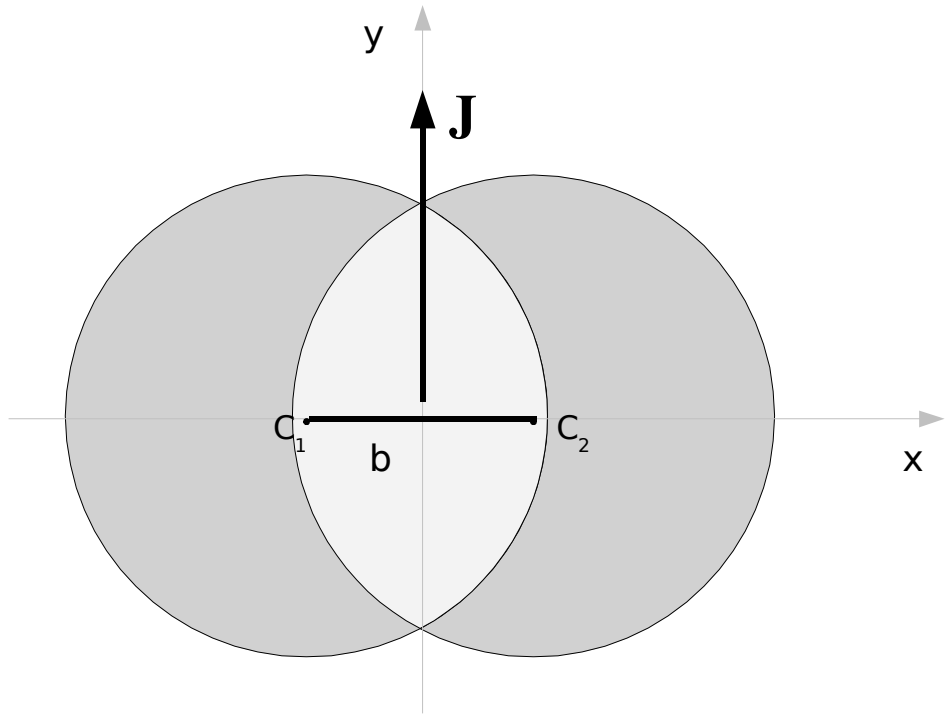}
  \caption{Sketch of a peripheral heavy ion collision at very high energy in
           the transverse projection. The overlap almond-shaped region is marked 
           in light grey and has an overall angular momentum directed along the 
           symmetry axis $y$, orthogonal to the reaction plane.}
\label{transverse}
\end{figure}
%-------------------------------------------------------------------------

\begin{figure}
  \includegraphics[height=.4\textheight]{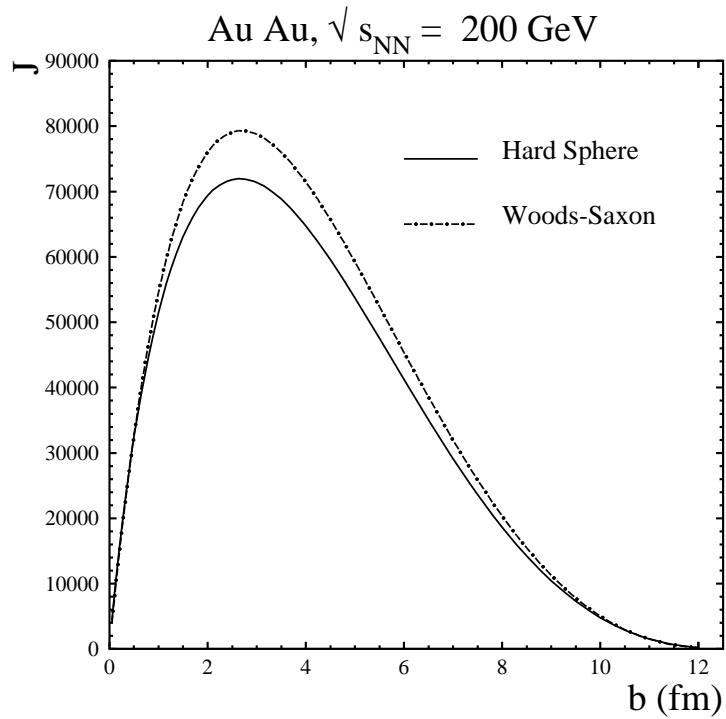}
  \caption{Angular momentum $J$ of the interaction region as a function of the
           impact parameter for Au-Au collisions at $\sqrt{s}_{NN}=200$ GeV.}
 \label{j}
\end{figure}

In the usual picture of a peripheral heavy ion collision at ultrarelativistic energy
the overlapping region of the two incoming nuclei gives rise to QGP whereas
the non-overlapping fragments fly away almost unaffected. Thereby, only a fraction
of the initial angular momentum $L_0$ is left to the interaction region, while the
largest part is carried away by the fragments (see fig.~\ref{collision}). The angular 
momentum of the interaction region takes its origin from the inhomogeneity of the 
density profile in the transverse plane, the so-called thickness function. This 
is much clearly seen in a longitudinal projection: the colliding strips of nucleons 
have, in peripheral collisions, different number 
of nucleons. While the central strips have the same weight, the strips above
it will have a net momentum directed along the negative $z$ axis and conversely
the ones below it (see fig.~\ref{collision}). The net momentum density at each point 
$(x,y)$ of the overlap region in the transverse plane (see fig.~\ref{transverse} 
for the axes definition) for symmetric (equal nuclei) collisions reads:
\begin{equation}\label{density}
  \frac{\d P}{\d x \d y} = [T(x-b/2,y)-T(x+b/2,y)]\frac{\sqrt{s}_{NN}}{2}
\end{equation}
where $T(x,y)$ is the thickness function, i.e. the longitudinal integral of the 
nucleon density:
$$
  T(x,y) = \int \d z \; n(x,y,z)
$$
Only if the two colliding objects were homogeneous in the transverse plane, the 
angular momentum of the interaction region would be vanishing. Yet, the nuclei 
{\em are not} homogenous in the transverse plane; for instance, if they are assumed 
to be homogenous spheres in their rest frame, their thickness function $T(x,y)$ would 
be proportional to $\sqrt{R^2-r^2}$, $r$ being the distance from the centre of 
the nucleus and $R$ its radius. In this case, eq.~(\ref{density}) would become:
\begin{equation}\label{density2}
  \frac{\d P}{\d x \d y} = 2n_0 \left[
  \sqrt{R^2-y^2-(x-b/2)^2}-\sqrt{R^2-y^2-(x+b/2)^2}\right]
  \frac{\sqrt{s}_{NN}}{2}
\end{equation}
From this momentum density, one gets an initial angular momentum $J$ of the 
interaction region directed along the $y$ axis:
\begin{equation}\label{angmom}
  {\bf J} = 2n_0 \int \d x \int \d y \; 
  x \left [ \sqrt{R^2-y^2-(x-b/2)^2}-\sqrt{R^2-y^2-(x+b/2)^2} \right]
  \frac{\sqrt{s}_{NN}}{2} {\bf \hat j} 
\end{equation}
In fig.~\ref{j} we show ${\bf J}$ for two colliding Gold nuclei at $\sqrt{s}_{NN} = 200$ GeV, 
in the two cases of hard spheres and Woods-Saxon distribution. For the former case,
it is seen that the angular momentum attains a maximal value at an impact parameter 
of 2.5 fm and quickly drops thereafter. The maximal value of $J$ is about 
$7.2 \times 10^4$, i.e. 29\% of the initial orbital angular momentum $L_0$ of the 
colliding nuclei at that impact parameter. Therefore, $J$ is very large and strongly 
dependent on the impact parameter $b$ but this effect is usually ignored in the initial 
conditions assumed for hydrodynamical calculations as in the commonly used Bjorken 
model the longitudinal flow velocity only depends on $z$ and it does not thus have any 
azimuthal anisotropy. 

%-------------------------------------------------------------------------
%-------------------------------------------------------------------------

This can be seen again from fig.~\ref{collision}; since the net momentum of the 
colliding strips varies monotonically along $x$, either the proper energy density 
or the fluid four-velocity or both must have an asymmetric profile in $x$ for the 
initial angular momentum to be conserved. If we take the reasonable assumption that
the proper energy density cannot have such an asymmetric dependence on $x$ because 
it can only depends on the density of nucleons at each point, the only remaining 
possibility is to admit that the initial longitudinal flow velocity is asymmetric 
in $x$ from the very beginning, i.e. it is azimuthally anisotropic, in such a way that:
\begin{equation}\label{jcons}
 - \int \d^3 \x \; x \, T^{0z} = - \int \d^3 \x \; x (\rho + p) \gamma^2 v_z(x)  
  = J  
\end{equation}
for a perfect fluid and if the initial flow transverse flow velocity is zero; 
in the above equation $T$ is the stress-energy tensor, $\rho$ is the proper energy 
density, $p$ the pressure and $\gamma^2 = (1-v^2)^{-1}$. The fact that $v_z$ is 
not azimuthally isotropicis implies, in general, a non-vanishing vorticity 
$\omegav=(1/2) \nabla \times {\bf v}$ for the fluid motion, and this may have 
remarkable consequences on the final particle spectra. It should be pointed out 
that some calculations \cite{hirano} indeed 
introduce an $x$-asymmetric proper energy density function. Still, even if $\rho$ 
was forced to have such an asymmetric $x$ dependence in order to fulfill angular
momentum conservation (\ref{jcons}), the final velocity field would not be the 
same as when, more reasonably, $v_z$ is asymmetric in $x$. We will try to illustrate 
such effects with an oversimplified hydrodynamical scheme in the next section.

%*****************************************************************************
\section{Hydrodynamical scheme}
%*****************************************************************************

We are now going to set up a very simple hydrodynamical scheme to show that 
the azimuthal anisotropy of the longitudinal flow velocity required by the angular
momentum conservation must enhance the elliptic flow.

\begin{figure}
  \includegraphics[height=.4\textheight]{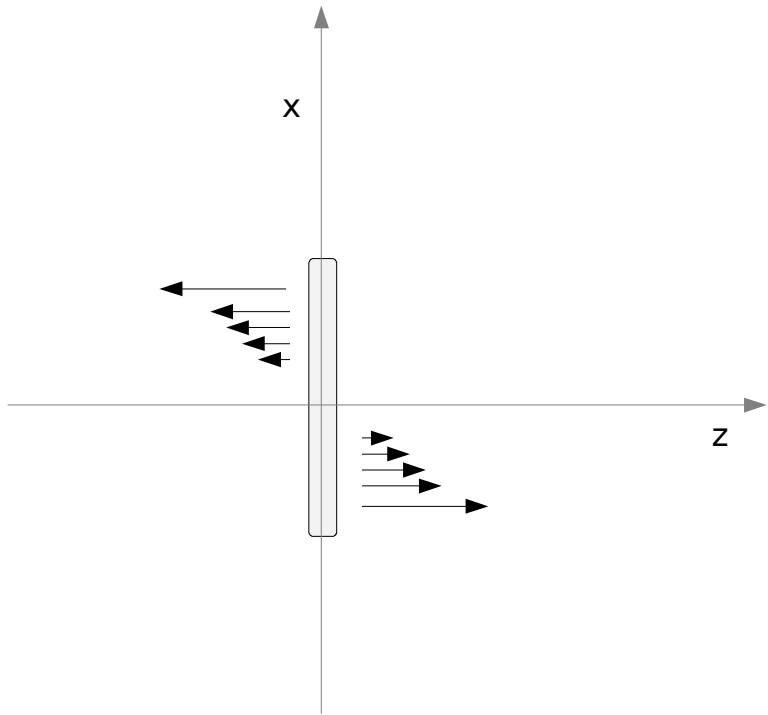}
  \caption{Initial longitudinal velocity profile for the limiting case of
           sudden thermalization in the very thin overlap region of the
           colliding ultrarelativistic nuclei.}
\label{slab}
\end{figure}

As has been mentioned, the requirement of an initial azimuthal anisotropy of 
the longitudinal flow velocity breaks the usual Bjorken scheme, where $v_z = z/t$.
As a first step, one would like to introduce a minimal change of the Bjorken
scheme, which is not an easy task though. Thus, to describe the possible
effects of an initial dependence of $v_z$ on $x$ in the most transparent way, 
we will assume an oversimplified scheme assuming that the two colliding nuclei 
give rise to a complete thermalization within an infinitesimally thin slab $\Delta z$ 
at the time $t=0$ (see fig.~\ref{slab}). This scheme looks very similar to Landau 
hydrodynamical model, were not for the inclusion of an initial flow velocity $v_z(x)$
which ought to vanish in $x=0$ for an infinitely thin slab, for evident symmetry 
reasons. Such a picture of the collision should be more realistic at asymptotically
large energies, where one expects thermalization to be extremely quick and nuclei
are infinitely Lorentz-contracted along their collision axis. Furthermore, we will 
assume to deal with a perfect fluid and we will focus our attention on the 
transverse motion only.

%-------------------------------------------------------------------------
%-------------------------------------------------------------------------

Firstly, we can write a relation for the initial momentum density 
\begin{equation}\label{momdens}
 (\rho_0 + p_0) \gamma_0^2 v_{z0} = \frac{1}{\Delta z} \frac{\d P}{\d x \d y} 
\end{equation}
where $\rho_0,p_0$ and $v_{z0}$ are the proper energy density, pressure and 
longitudinal flow velocity at the time $t=0$, with $\gamma_0^2=1/(1-v_{z0}^2)$
because the initial transverse velocity is vanishing; $\d P/\d x \d y$ is given 
by eq.~(\ref{density}). The eq.~(\ref{momdens})
makes it clear that $v_{z0}$ in this approach cannot be zero because of the 
initial unbalance in momentum and that, in general, depends on both $x$ and $y$. The
specific functional form of the proper energy density $\rho$ affects the functional 
dependence of $v_{z0}$ but it should not suppress its dependence on $x$ because 
it is reasonable to assume that it has a symmetric dependence on $x$, unlike 
$\d P /\d x \d y$, as pointed out at the end of Sect.~2. From a hydrodynamical 
point of view, the remarkable consequence of this is that the initial vorticity 
$\omegav = (1/2) \nabla \times {\bf v}$ is non-vanishing, unlike in the traditional 
Bjorken picture where $v_z$ only depends on $z$. Indeed:
\begin{equation}\label{vort}
 \omega_{x}(t=0) =  \frac{1}{2}\frac{\partial v_{z0}}{\partial y} \qquad 
 \omega_{y}(t=0) = -\frac{1}{2}\frac{\partial v_{z0}}{\partial x} 
\end{equation}
where the largest component among the two is the one along $y$ axis, because of the
asymmetry of $v_{z0}$ with respect to $y$ axis. Conversely, $v_{z0}$ is {\em symmetric}
with respect to the $x$ axis (see eqs.~(\ref{density}),(\ref{density2})) and its 
partial derivative with respect to $y$ ought to vanish in $y=0$ for any $x$. 

It is worth pointing out that the evolution equation for the classical vorticity 
$\omegav$ in the relativistic case is more complicated than in non-relativistic 
fluid mechanics. However, it is still true that a non-vanishing initial value of 
the classical vorticity makes the fluid motion a vorticous one in general.
This stems from the Carter-Lichnerowicz equation of motion (equivalent to the
Euler equations) for a perfect fluid with one conserved charge \cite{gourgo}:
\begin{equation}\label{cl}
 u^{\mu} \left(\partial_{\mu} \bar h u_\nu - \partial_\nu \bar h u_\mu \right)
 = T \partial_\nu \bar s
\end{equation}
where $\bar h = (\rho + p)/n$ and $\bar s = s/n$ are the enthalpy and the entropy 
densities normalized to the charge density $n$, $p$ is the pressure and $T$ is the 
temperature. The {\em vorticity tensor} is usually defined as:
\begin{equation}\label{vortens}
 \Omega_{\mu \nu} = 
  \left(\partial_{\mu} \bar h u_\nu - \partial_\nu \bar h u_\mu \right) 
\end{equation}
and the {\em vorticity vector} as \cite{gourgo}:
\begin{equation}\label{vortvect}
 \omega^\alpha = \frac{1}{4 \bar h} \epsilon^{\alpha \mu \nu \sigma}
 u_\mu \Omega_{\nu \sigma} = \frac{1}{4} \epsilon^{\alpha \mu \nu \sigma}
 u_\mu \left(\partial_{\nu} u_\sigma - \partial_\sigma u_\nu \right) 
\end{equation}
It is quite straightforward to show from eq.~(\ref{vortvect}) that the vorticity 
vector field has the non-relativistic limit:
\begin{equation}
 \omega \rightarrow \left( 0, \frac{1}{2} \nabla \times {\bf v} \right)
\end{equation}
so that $\omega$ is a proper relativistic generalization of the classical vorticity
$\omegav = (1/2) \nabla \times {\bf v}$. The time component of $\omega$ is:
$$
 \omega^0 = \frac{1}{2} \gamma^2 {\bf v} \cdot \nabla \times {\bf v}
$$
Thus, if $\nabla \times {\bf v} \ne 0$ then $\omega \ne 0$ and $\Omega \ne 0$. 
It can be shown, starting from (\ref{cl}), that the spacial part of the vorticity 
tensor, that is 
$\Omegav = \nabla \times \bar h \gamma {\bf v}$, fulfills the Helmholtz vorticity 
equation:
\begin{equation}
 \frac{\partial \Omegav}{\partial t} = \nabla \times ({\bf v} \times \Omegav )
\end{equation}
provided that the fluid is {\em isentropic}, i.e. $\nabla \bar s =0$. All classical
consequences of the vorticity equations then hold in relativity provided that 
$\omegav$ is replaced by $\Omegav$.

Let us now study more in detail the fluid equations of motion at the time $t=0$.
We will write the Euler equation, instead of the Carter-Lichnerowicz, for a perfect
ultrarelativistic fluid, with equation of state $p=\rho/3$. Accordingly:
\begin{equation}
  (\rho + p ) (u \cdot \partial) u^\mu = g^{\mu \nu} \partial_\nu p - 
   (u \cdot \partial p) u^\mu 
\end{equation}
and focus on the transverse components at the time $t=0$, when $u_x=u_y=0$, 
namely:
\begin{equation}\label{init1}
   \rho_0 \gamma_0 \frac{\partial u_i}{\partial t} \Big|_{t=0} = 
 - \frac{1}{4}\frac{\partial \rho}{\partial x_i} \Big|_{t=0}
\end{equation}
for $i=1,2$ where the equation of state has been used. Multiplying both sides by
$\gamma_0^2$ and manipulating the derivative on the right hand side:
\begin{equation}\label{init2}
   \rho_0 \gamma_0^3 \frac{\partial u_i}{\partial t} \Big|_{t=0} = 
 - \frac{1}{4}\frac{\partial \rho \gamma^2}{\partial x_i} \Big|_{t=0}
 + \frac{1}{4} \rho_0 \frac{\partial \gamma^2}{\partial x_i} \Big|_{t=0}
\end{equation}
Now, since $\gamma_0^2 = 1/(1-v_{z0}^2)$:
\begin{equation}
 \frac{\partial \gamma}{\partial x_i}\Big|_{t=0} = \gamma_0^3 v_{z0} 
  \frac{\partial v_{z0}}{\partial x_i}
\end{equation}
and eq.~(\ref{init2}) becomes:
\begin{equation}\label{init3}
   \rho_0 \gamma_0^3 \frac{\partial u_i}{\partial t} \Big|_{t=0} = 
 - \frac{1}{4}\frac{\partial \rho \gamma^2}{\partial x_i} \Big|_{t=0}
 + \frac{1}{4} 2 \rho_0 \gamma_0^4 v_{z0} 
   \frac{\partial v_{z0}}{\partial x_i} \Big|_{t=0}
\end{equation}
Two terms are then responsible for the initial transverse velocity increase: the 
first is related to the gradient of the energy density in the observer frame, 
while the second depends on the gradient of the initial velocity field, i.e. on the 
vorticity according to eq.~(\ref{vort}). If $v_{z0}$ was independent of $x,y$
the second term would vanish and the transverse expansion would then be driven
by the energy density gradient only, like in the usual picture. Because of the 
eccentrity of the overlap region 
(see fig.~\ref{transverse}), the system gets an initial kick larger in the $x$ direction
than in $y$ and an anisotropy in the final spectra ensues. If, however, the second
term is included, the expansion gets an additional contribution because 
$\partial v_{z0}/\partial x < 0$ (see fig.~\ref{slab}), and $v_{z0}$ is negative 
for $x>0$ and positive for $x<0$ so that altogether the second term drives an increase
of $u_x$ for $x>0$ and a decrease (starting from zero) for $x<0$. Moreover, the 
expansion rate related to this term will be larger in the $x$ direction than in $y$ 
because, expectedly, $\partial v_{z0}/\partial x > \partial v_{z0}/\partial y$,
(see discussion following eq.~(\ref{vort})), thereby enhancing the elliptic flow. In other 
words, besides the {\em geometrical
anisotropy}, elliptic flow gets a finite contribution from an initial 
{\em kinematical anisotropy} of the longitudinal velocity. 
This enhancement of elliptic flow can be seen as a centrifugal effect owing to 
angular momentum conservation: particles with a momentum orthogonal to {\bf J},
i.e. directed along the reaction plane get an additional momentum kick with 
respect to those emitted along {\bf J}.

%-------------------------------------------------------------------------
\begin{figure}
  \includegraphics[height=.4\textheight]{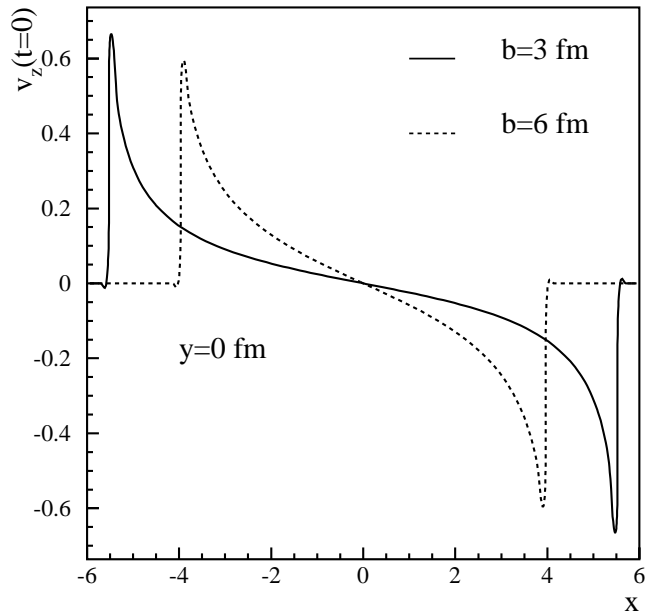}
  \caption{Initial longitudinal velocity profile along the reaction plane 
  $y=0$ for two different impact parameters for the collision of two hard 
  sphere nuclei with radius 7 fm.}
 \label{vz}
\end{figure}
%-------------------------------------------------------------------------

\begin{figure}
  \includegraphics[height=.4\textheight]{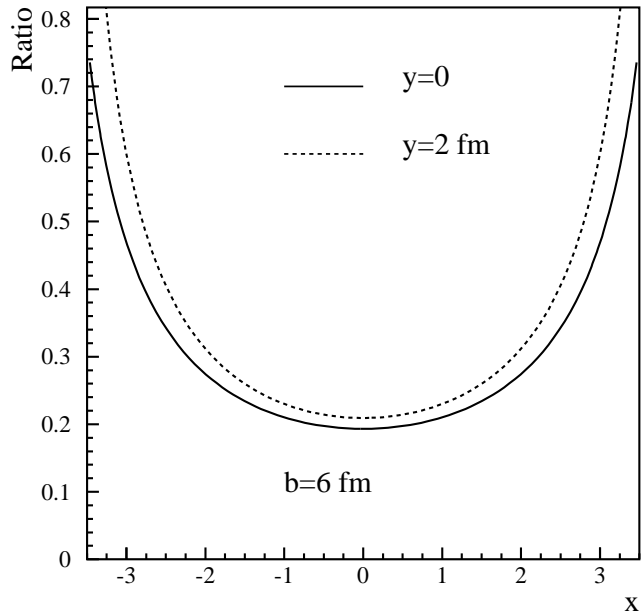}
  \caption{Ratio of the term proportional to the vorticity and
   the term proportional to energy density gradient along $x$ in eq.~(\ref{init3})
   as a function of $x$ for $y=0$ and $y=2$ fm for the collision of two hard sphere
   nuclei with radius 7 fm at an impact parameter $b=6$ fm.}
 \label{ratio}
\end{figure}

It is now interesting to make an estimate of how large this contribution is in our 
simple scheme. Assuming that the energy density is proportional to the total energy 
of nucleons in the overlap region so that:
\begin{equation}\label{endens}
   (\rho_0 + p_0) \gamma^2_0 - p_0 = 
  \frac{1}{\Delta z} \frac{\d E}{\d x \d y} = \frac{1}{\Delta z}
  [T(x-b/2,y)+T(x+b/2,y)]\frac{\sqrt{s}_{NN}}{2}
\end{equation}
and, by using (\ref{momdens}) and the equation of state $p=\rho/3$ we can obtain the 
expressions of the initial proper energy density $\rho$:
\begin{equation}\label{rho}
   \rho_0 = \frac{1}{\Delta z} \sqrt{4 \left(\frac{\d E}{\d x \d y}\right)^2
                - 3 \left(\frac{\d P}{\d x \d y}\right)^2} -\frac{1}{\Delta z}
                \frac{\d E}{\d x \d y}
\end{equation}
and the flow velocity $v_{z0}$:
\begin{equation}\label{flowvel}
   v_{z0} 
 =  \frac{3\frac{\d P}{\d x \d y}}{\sqrt{4 \left(\frac{\d E}{\d x \d y}\right)^2
   - 3 \left(\frac{\d P}{\d x \d y}\right)^2} + 2 \frac{\d E}{\d x \d y}} 
\end{equation}
which is shown in fig.~\ref{vz} for the case of hard sphere nuclei with radius
7 fm. According to eq.~(rho), the proper energy density is an even function of
$x$, as it was expected with the assumption (\ref{endens}), while $v_{z0}$ is an
odd function of $x$. Also, tt can be seen from fig.~\ref{vz} that $v_{z0}$ has 
a singular derivative at the edge of the 
overlap region which is due to the hard sphere assumption; such singularities 
disappear with smooth density profiles. By using (\ref{flowvel}), (\ref{rho}),
(\ref{momdens}) and (\ref{endens}) we can compute the ratio of the second to the 
first term in eq.~(\ref{init3}) for the $x$ axis:
\begin{equation}\label{ratio0}
  -\frac{ 2 \rho_0 \gamma^4_0 v_{z0} \frac{\partial v_{z0}}{\partial x}\Big|_{t=0}}
  {\frac{\partial \rho \gamma^2}{\partial x}\Big|_{t=0}}
\end{equation}
in order to evaluate the importance of the vorticity term for the expansion rate.
This ratio is shown in fig.~\ref{ratio} for the case of hard sphere nuclei 
for two different $y$ values at an impact parameter $b=6$ fm. It is seen that the second 
term is a consistent fraction of the first term even near the collisions centre $x=0$
(about 20\%) while it steeply increases at larger $x$ values; at the boundary of the 
$x$ interval the ratio shows spikes owing to the sharp sphere assumption and it is 
not shown.
Of course, these numbers refer to an oversimplified example and just for the initial
expansion kick, but the conclusion that the longitudinal velocity gradient cannot be 
neglected in more realistic hydrodynamical calculations should hold. 

%-------------------------------------------------------------------------
%-------------------------------------------------------------------------

As has been mentioned, in some hydrodynamical calculations \cite{hirano,heinzkolb}, 
a non-vanishing angular momentum of the plasma is tacitly introduced by enforcing 
an asymmetric $x$ dependence for the proper energy density in peripheral collisions
keeping the Bjorken longitudinal scaling, i.e. the independence of $v_z$ on the
coordinates $x,y$. Thereby, the longitudinal momentum density (\ref{momdens}) 
conservation is fulfilled even though $v_z$ is independent of $x$ and the angular
momentum conservation (\ref{jcons}) is also fulfilled. We think that this assumption
is quite unnatural. Firstly, it cannot hold in our specific example of instantaneous 
thermalization at infinitely large energy (with infinitesimally thin fluid in 
fig.~\ref{slab}) because the only velocity which is compatible with symmetry and 
independent of $x$ is 0, thus making both momentum and angular momentum density 
vanishing. However, even in the more realistic and more general case of finite 
thermalization time, it does not lead to the same flow velocity field as in the 
case of angular momentum conserved through Bjorken scaling breaking because of the 
{\em absence of the vorticity term}. This can be shown by enforcing the equality 
of angular momentum densities in the two approaches:
\begin{equation}\label{equality}
   \frac{4}{3} \tilde \rho_0 \tilde \gamma_0^2 \tilde v_{z0} = 
   \frac{4}{3} \rho_0 \gamma_0^2 v_{z0}
\end{equation}
where quantities with a tilde on the left hand side are such that only $\tilde\rho$
depends on $x$ while on the right hand side we have the standard ones in our 
approach. From the above equation follows:
\begin{equation}\label{interm}
 \frac{\partial \tilde \rho}{\partial x}\Big|_{t=0} \tilde \gamma_0^2 \tilde v_{z0}
 = \frac{\partial \rho}{\partial x}\Big|_{t=0} \gamma_0^2 v_{z0} + 
  \rho_0 \frac{\partial \gamma_0^2 v_{z0}}{\partial x}\Big|_{t=0} 
\end{equation}
Using (\ref{interm}) and (\ref{equality}) to obtain $\partial \rho/ \partial x$ in 
the equation of motion at the time $t=0$ (\ref{init1}), we get, after some manipulations:
\begin{equation}\label{init4}
   \frac{\partial u_x}{\partial t}\Big|_{t=0} = -\frac{1}{4 \gamma_0 \rho_0}
   \frac{\partial \rho}{\partial x}\Big|_{t=0} = - \frac{1}{4 \tilde \rho_0 \tilde \gamma_0} 
   \frac{\partial \tilde \rho}{\partial x}\Big|_{t=0} \frac{\tilde \gamma_0}{\gamma_0} 
   + \frac{1}{4 \gamma_0^3 v_{z0}}\frac{\partial \gamma^2 v_{z0}}{\partial x}\Big|_{t=0}   
\end{equation}
Conversely, in an approach where velocity is uniform with the same angular momentum
density, one would have:
\begin{equation}\label{initf}
   \frac{\partial u_x}{\partial t} \Big|_{t=0} = - \frac{1}{4 \tilde \rho_0 \tilde 
   \gamma_0} \frac{\partial \tilde \rho}{\partial x} \Big|_{t=0} 
\end{equation}
Therefore, even though the same angular momentum density was enforced by modifying
the energy density profile, the expansion rate could be consistently different from 
the one with non-vanishing vorticity because of the factor $\tilde\gamma/\gamma$ and,
chiefly, the additional term proportional to the derivative of longitudinal velocity
which in general speeds up expansion, as we have seen.

The ratio of the expansion-driving terms in eq.~(\ref{ratio0}) depends only on geometry 
and not on the centre-of-mass energy, because so do both $\rho$ and $v_{z0}$ according 
to eqs.~(\ref{rho}),(\ref{flowvel}) and the expressions (\ref{momdens}),(\ref{endens}). 
This apparent energy-independence is just a specific feature of our simple scheme 
where the longitudinal dimension was shrunk to a very thin slab. In fact, this cannot 
be the case in a more realistic one, where thermalization is not instantaneous and, 
therefore, the gradient of $v_z$ is distributed on a larger volume. In other words, 
the relative vorticity contribution will not be as large as it turned out to be by 
enforcing instantaneous thermalization in an infinitesimal slab $\Delta z$. This can
be better seen by rewriting the angular momentum, for a perfect fluid, as:
\begin{eqnarray}\label{jvort}
{\bf J} &=& \int \d^3 \x \; {\bf x} \times \piv = \int \d^3 \x \; \nabla \frac{\x^2}{2} 
 \times h \gamma^2 {\bf v} = \int \d^3 \x \; \nabla \times \frac{\x^2}{2} h \gamma^2 {\bf v} -
  \int \d^3 \x \; \frac{\x^2}{2} \nabla \times h \gamma^2 {\bf v} \nonumber \\
&=&  - \int \d^3 \x \; \frac{\x^2}{2} \nabla h \gamma^2 \times {\bf v} -
       \int \d^3 \x \; \x^2 h \gamma^2 \omegav 
\end{eqnarray}
where $\pi_i = T^{0i} = h \gamma^2 v_i$ is the momentum density, $h = \rho + p$
is the enthalpy density and $\omegav = (1/2) \nabla \times {\bf v}$; in the above 
equation we assumed that the enthalpy density vanishes outside a compact region.
In general, the sum of the two terms in eq.~(\ref{jvort}) is constrained by angular 
momentum conservation but their relative contribution to it can, and will, vary 
with the centre-of-mass energy. Most likely, at lower energy, the vorticity term 
will be less important whereas, at higher energy, its relative contribution should 
approach the limiting one calculated in our simple scheme because thermalization 
is expected to be faster, with an initial denser plasma and a higher angular momentum
density. If this is the case, at the 
LHC, a further increase of the elliptic flow with respect to RHIC ought to be observed.

%-------------------------------------------------------------------------
\begin{figure}
  \includegraphics[height=.5\textheight]{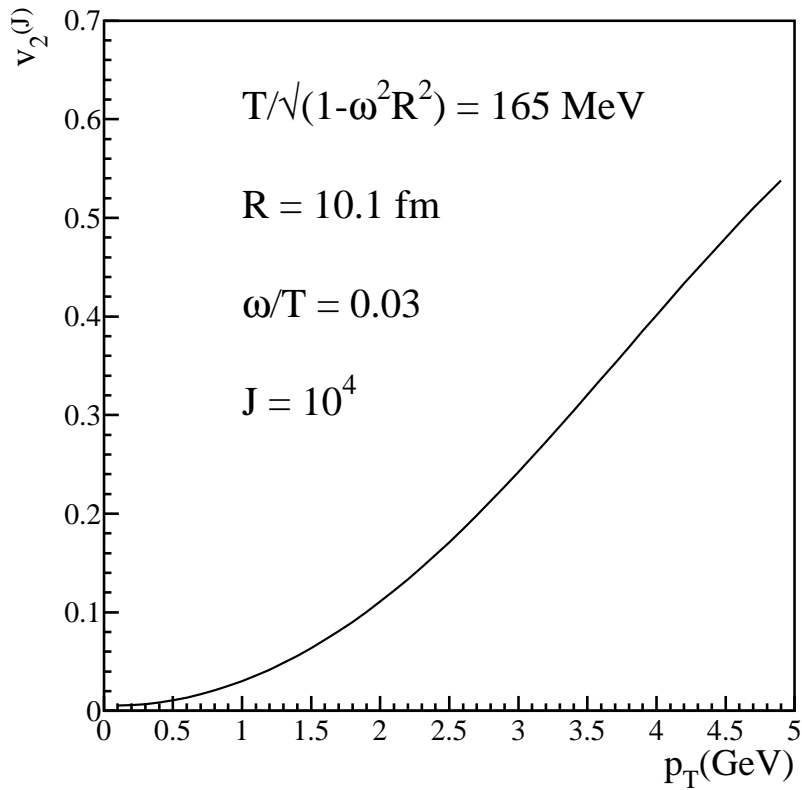}
  \caption{Elliptic flow coefficient $v_2^{(J)}$ as a function of $p_T$ for  
           hadrons originated from a spherical spinning plasma at a chemical 
           freeze-out T=165 MeV and radius 10.1 fm for $\omega/T=0.03$. The
           elliptic flow would simply vanish if $J=0$.}
 \label{v2}
\end{figure}
%-------------------------------------------------------------------------

%*****************************************************************************
\section{Elliptic flow for a spinning system}
%*****************************************************************************

Taking viscosity into account implies a modification of the hydrodynamical equations
but this should not affect our conclusions. This can be understood with a simple 
argument: angular momentum has to be conserved anyway and dissipative effects such 
as viscosity will speed up entropy increase. Thus, the system will tend to the 
maximal entropy configuration which, for a system with finite angular momentum and 
finite volume, is a rigidly spinning fluid, with velocity field ${\bf v} = \omegav 
\times {\bf x}$, $\omegav$ being a constant vector related to the total angular 
momentum ${\bf J}$ \cite{landau,becapicc}. Of course, the quick expansion will prevent 
the system from reaching global equilibrium before decoupling, still the system will 
evolve towards that configuration. 

Hence, although viscosity, along with other dissipative forces, will not be able to 
create a fully equilibrated rigidly spinning plasma fireball, it is interesting to 
show that in this ideal case the effect of a finite angular momentum on elliptic 
flow and other observables is remarkable. 

That a globally spinning interaction region brings about an anisotropy in the particle 
azimuthal spectra was argued many years ago by Hagedorn \cite{hagewam} and recently 
rediscussed in ref.~\cite{satz}. Assuming statistical hadronization for a fully 
equilibrated subsystem of the plasma, the elliptic flow coefficient can be calculated
as that of a rigidly spinning ideal hadron-resonance gas with angular momentum 
${\bf J}_\omega$ such that $J_\omega < J$ and fixed angular velocity $\omegav = (1/2) 
\nabla \times {\bf v}$ parallel to it and linked to ${\bf J}_\omega$ by a thermodynamic 
relation which is linear for small $\omega/T$ values \cite{becapicc}. In the Boltzmann 
approximation for primary hadrons, this reads~\cite{becapicc}:
\begin{equation}\label{v2j}
 v_2^{(J)} = 
 \frac{\int {\rm d}^3 x \frac{{\rm K}_1(m_T\sqrt{1-\vert{\bf \omega} \times 
 {\bf x}\vert^2_{\Vert}}/T)}{\sqrt{1-\vert{\bf \omega} \times {\bf x}\vert^2_{\Vert}}}
 {\rm I}_2\left( \frac{p_T z \omega}{T}\right)}
 {\int {\rm d}^3 x \frac{{\rm K}_1(m_T\sqrt{1-\vert{\bf \omega} \times {\bf x}
 \vert^2_{\Vert}}/T)}{\sqrt{1-\vert{\bf \omega} \times {\bf x}\vert^2_{\Vert}}}
 {\rm I}_0\left( \frac{p_T z \omega}{T}\right)}
\end{equation}
where ${\rm K}_1$, ${\rm I}_n$ are modified Bessel functions and $T$ is the global
temperature. It should be stressed that the global temperature $T$ in a spinning 
relativistic gas is related to the {\em local} proper temperature $T_0$, measured
by a comoving thermometer, by the relation \cite{becapicc}:
\begin{equation}\label{local}
  T_0(r) = \frac{T}{\sqrt{1 - \omega^2 r^2}}
\end{equation}
where $r$ is the distance from the rotation axis $\omegav$. Since it is the local, 
and not the global, temperature which determines the phase of the system, the decoupling 
should occur when the highest local temperature reaches the critical value $T_c$ for 
the quark-hadron transition, that is when:
\begin{equation}
  \frac{T}{\sqrt{1 - \omega^2 R^2}} = T_c
\end{equation}
being $R$ the maximal distance from the rotation axis. 

The behaviour of the ``rotational" $v_2^{(J)}$ (which would simply vanish if $J=0$)
as a function of $p_T$ for primary hadrons is very similar to that driven by pressure 
gradients in usual hydrodynamical
calculations, and it turns out to be almost independent of the particle mass. It is
shown in fig.~\ref{v2} for $\omega/T = 0.03$ at the chemical freeze-out temperature $T_c=165$ 
MeV, for a spherical source with radius $R=10.1$ fm, for a total angular momentum 
$J_\omega \simeq 10^4$, i.e. of the same order of the $J$ of the interaction region
at RHIC energies. The $v_2^{(J)}$ of primary hadrons from a globally spinning region 
would be therefore very large, although resonance decays should lower the final one 
consistently. 

%*****************************************************************************
\section{Polarization}
%*****************************************************************************

Elliptic flow is not a unique consequence of an intrinsic rotation. 
There is, however, a distinctive signature thereof: a polarization of the emitted 
hadrons along the angular momentum direction (in the observer frame). That a
large angular momentum in peripheral heavy ion collisions could give rise to
polarization of the final hadrons has been first proposed in refs.~\cite{xnwang}, 
where a quantitative assessment was performed within a perturbative QCD framework,
with the polarization of quarks assumed to be effectively transferred to final 
hadrons. Recently, it has been pointed out that a plasma with polarized quarks
could be probed by observing the polarization of direct photons \cite{ipp}.
We take a different approach here and we determine the polarization of particles 
invoking local thermodynamical equilibrium and the statistical hadronization dogma 
which is succesfull in describing hadronic multiplicities: every multihadronic 
state compatible with conservation laws is equally likely. Therefore, since the 
total angular momentum is not vanishing, when the plasma hadronizes, available spin 
states will not be evenly populated and a net polarization of the produced hadrons 
will show up. In this approach, there is no need to invoke any special dynamical 
mechanism for the polarization of quarks to be transferred to hadrons, as it 
should happen as a consequence of the statistical nature of this process.

The proper polarization vector $\Piv_0$ of particles in a relativistic rotating 
ideal gas has been calculated by the authors \cite{becapicc}:
\begin{equation}
\Piv_0 = \frac{1}{2} \tanh \frac{\omega}{2T} 
 \left[ \frac{\varepsilon}{m} \hat \omegav - \frac{\hat \omegav \cdot {\bf p} 
 {\bf p}}{m(\varepsilon+m)} \right] 
\end{equation}
for spin 1/2 particles and:
\begin{equation}\label{anyspin}
 \Piv_0 = \frac{\sum_{n=-S}^S n \e^{n\omega/T}}{\sum_{n=-S}^S \e^{n\omega/T}} 
  \left[ \frac{\varepsilon}{m} \hat \omegav - \frac{\hat \omegav \cdot {\bf p} 
 {\bf p}}{m(\varepsilon+m)} 
 \right]
\end{equation}
for generic spin particles, where $\varepsilon$ is the energy and ${\bf p}$ the 
momentum of the particle. The polarization along $\hat\omegav$, i.e. $\Piv_0 \cdot
\hat\omegav$ turns out to be maximal for particles emitted orthogonally to 
$\omegav$ (i.e. along the reaction plane for an equilibrated spinning system) 
and increases for increasing $p_T$ up to momenta of the order of $2 m T/\omega$ 
where the rotational grand-canonical ensemble scheme fails and more complicated 
expression arise \cite{becapicc}. Also, the vector mesons show spin alignment in 
that the $00$ component of the spin density matrix turns out to be different 
from 1/3 and specifically \cite{becapicc}: 
\begin{equation}\label{align}
  \rho_{\omega\,00}(p) = \frac{1}{2 \cosh (\omega/T) + 1}
 \left[ \cosh(\omega/T) + \frac{({\bf p} \cdot \omegav)^2}{\p^2 \omega^2} 
 \left( 1 - \cosh(\omega/T) \right) \right]
\end{equation}
which, for small $\omega/T$, reduces to:
\begin{equation}
  \rho_{\omega\,00}(p) \simeq \frac{1}{3} + 
  \frac{1-3({\hat {\bf p}}\cdot{\hat \omegav})^2}{18} \frac{\omega^2}{T^2}
\end{equation}

It is interesting to note that the polarization (more generally the spin density
matrix) depends on the ratio between angular velocity and global temperature, 
that is, using (\ref{local}) on $\gamma \omega/T_0$, being $T_0$ the local 
temperature. Therefore, it can be conjectured, by invoking locality principle, that 
a polarization should appear in a generic accelerated hydrodynamical cell at local 
equilibrium fully determined by local quantities. Hence,$\omegav$ is to be 
plausibly replaced by the vector:
\begin{equation}\label{locome}
  \omegav \rightarrow \frac{1}{v^2}{\bf v} \times {\bf a}
\end{equation}
which is the local angular velocity for a general trajectory, according to the
Frenet formulae. If this conjecture is true, every hadronizing hydrodynamical cell
will produce hadrons with polarization vector (\ref{anyspin}) with $\omegav$ equal
to the right hand side of (\ref{locome}).

The expected polarization values are of the order of $\omega/T$, which is reasonbly
some percent or less (see previous section) but they should increase with particle 
momenta up to momenta of few GeV's and hopefully become observable \footnote{Recent
measurements by RHIC experiments set a limit on average $\Lambda$ polarization to
be $\sim 0.02$ \cite{star}}. However, the expressions (\ref{anyspin}) and (\ref{align}) 
refer to primary hadrons, i.e. those emitted from the source at decoupling and resonance 
decays can further dilute the polarization, so that a more detailed study is needed.

It is difficult to predict the evolution of polarization values as a function
of centre-of-mass energy. However, it can be argued that they should increase by
considering that angular momentum {\em density} at freeze-out increases as a 
function of $\sqrt{s}_{NN}$. This happens because the size of the system at freeze-out
increases approximately logarithmically whereas the angular momentum of the 
interaction region increases linearly with $\sqrt{s}_{NN}$ (see eq.~(\ref{angmom})). 
Since the angular momentum density must be somehow related to the final local angular 
velocity (\ref{locome}), a fair conclusion follows that polarization effects 
should increase with the collision energy.

%*****************************************************************************
\section{Summary and conclusions}
%*****************************************************************************

We have pointed out that angular momentum conservation in peripheral ultrarelativistic 
heavy ion collisions at very high energy should give an additional contribution to the
azimuthal momentum anisotropy, thereby enhancing the elliptic flow coefficient $v_2$.
By using a very simple hydrodynamical scheme, we have shown that taking angular 
momentum conservation properly into account implies, most likely, a non-uniform 
longitudinal flow velocity in the transverse plane breaking the usual assumption of
Bjorken scaling. This in turn generates a non-vanishing initial vorticity term in 
the equations of motion which enhances the transverse expansion rate and may be 
able to balance the elliptic flow deficit observed by Heinz and Song \cite{hsong} 
in minimally viscous hydrodynamical calculations. Angular momentum conservation is 
also implemented in current hydrodynamical calculations by keeping the Bjorken
scaling hypothesis, but the resulting expansion rates are different. We expect 
this effect to be more visible at very large energies where the vorticity contribution 
to the angular momentum density tends to an upper geometrical limit, that we have 
analyzed in our simplified scheme. Hence, we predict that $v_2$ should increase 
from RHIC to LHC energy, although we cannot give a definite quantitative estimate. 

The most characteristic signature of the vorticity induced by angular momentum 
conservation would be a polarization of the emitted particles, which is predicted to be,
in the observer frame, for a globally spinning system, orthogonal to the reaction plane 
and maximal for particles with momentum parallel to the reaction plane in case of 
a globally spinning plasma. A quantitative assessment of these effects for the actual 
hydrodyamical evolution is very difficult, but we argued that the polarization should 
be there for a general accelerated fluid motion. Also this effect should be better
observed at the LHC, where the angular momentum density should be larger.

%*****************************************************************************
\section*{Acknowledgements}
%*****************************************************************************

We thank for very useful discussions U.~Heinz, R.~Jaffe, L.~Maiani, A.~Polosa, 
K.~Rajagopal, H.~Satz and X.~N.~Wang. We are grateful to G.~Torrieri for suggesting 
references in vorticous relativistic hydrodynamics. We thank Galileo Galilei Insitute 
for kind hospitality.

%*********************************************************


\begin{thebibliography}{99}
%*********************************************************

\bibitem{review} 
  P.~F.~Kolb and U.~W.~Heinz,
  ``Hydrodynamic description of ultrarelativistic heavy-ion collisions,''
  arXiv:nucl-th/0305084.

\bibitem{heinz}
  P.~F.~Kolb, J.~Sollfrank and U.~W.~Heinz, Phys.\ Rev.\  C {\bf 62}, 
  054909 (2000).

\bibitem{hirano}
  T.~Hirano and K.~Tsuda,  Phys.\ Rev.\  C {\bf 66}, 054905 (2002).

\bibitem{romat}
  P.~Romatschke and U.~Romatschke, Phys. Rev. Lett. 99, 172301 (2008).

\bibitem{hsong}
  H.~Song and U.~W.~Heinz, Phys. Lett. B 658, 279 (2008).
\bibitem{xnwang} 
  Z.~T.~Liang and X.~N.~Wang, Phys. Rev. Lett.  
 {\bf 94}, 102301 (2005).

\bibitem{wang}
  J.~H.~Gao, S.~W.~Chen, W.~t.~Deng, Z.~T.~Liang, Q.~Wang and X.~N.~Wang,
  ``Global quark polarization in non-central $A+A$ collisions,''
  arXiv:0710.2943.

\bibitem{torrieri} 
  B.~Betz, M.~Gyulassy and G.~Torrieri, Phys.\ Rev.\  C {\bf 76}, 044901 (2007).

\bibitem{becapicc} F.~Becattini, F.~Piccinini, ``Polarization and spectra
 in the ideal relativistic spinning gas", arXiv:0710.5694, to appear in Ann. Phys.

\bibitem{gourgo} E.~Gourgoulhon, ``An introduction to relativistic hydrodynamics",
 in  {\em Stellar Fluid Dynamics and Numerical Simulations: from the Sun to Neutron Stars}
 EAS Series {\bf 21} 43 (2006).

\bibitem{heinzkolb}
  U.~W.~Heinz and P.~F.~Kolb, J.\ Phys.\ G {\bf 30}, S1229 (2004)

\bibitem{landau} L.~D.~Landau and E.~M.~Lifshitz, ``Statistical Physics", 
 Pergamon Press, 1980.

\bibitem{hagewam} 
 R.~Hagedorn and U.~Wambach, Nucl.\ Phys.\  B {\bf 123}, 382 (1977).

\bibitem{satz} P.~Castorina, D.~Kharzeev, H.~Satz, Eur. Phys. J. C
{\bf 52} 187 (2007).

\bibitem{ipp} A.~Ipp, A.~Di Piazza, J.~Evers and C.~H.~Keitel, 
  ``Photon polarization as a probe for quark-gluon plasma dynamics,'' 
   arXiv:0710.5700.

\bibitem{star}  B.~I.~Abelev {\it et al.}  [STAR Collaboration],
  Phys.\ Rev.\  C {\bf 76}, 024915 (2007)

\end{thebibliography}
\end{document}